\documentclass[]{aastex631}

\shorttitle{Chiron's ``Rings'' in 2018}
\shortauthors{Sickafoose et al.}

\begin{document}

\title{Material Around the Centaur (2060) Chiron from the 2018 November 28 UT Stellar Occultation}

\correspondingauthor{A. Sickafoose}
\email{asickafoose@psi.edu}

\author[0000-0002-9468-7477]{Amanda A. Sickafoose}
\affiliation{Planetary Science Institute \\
1700 East Fort Lowell, Suite 106 \\
Tucson, AZ 85719, USA}

\author{Stephen E. Levine}
\affiliation{Lowell Observatory \\
1400 West Mars Hill Road \\
Flagstaff, AZ 86001, USA}
\affiliation{Department of Earth, Atmospheric, and Planetary Sciences \\
Massachusetts Institute of Technology \\
77 Massachusetts Ave. \\
Cambridge, MA 02139, USA}

\author{Amanda S. Bosh}
\affiliation{Lowell Observatory \\
1400 West Mars Hill Road \\
Flagstaff, AZ 86001, USA}

\author{Michael J. Person}
\affiliation{Department of Earth, Atmospheric, and Planetary Sciences \\
Massachusetts Institute of Technology \\
77 Massachusetts Ave. \\
Cambridge, MA 02139, USA}

\author{Carlos A. Zuluaga}
\affiliation{Department of Earth, Atmospheric, and Planetary Sciences \\
Massachusetts Institute of Technology \\
77 Massachusetts Ave. \\
Cambridge, MA 02139, USA}

\author{Bastian Knieling}
\affiliation{Deutsches SOFIA Institut\\
Universit\"at Stuttgart\\
Pfaffenwaldring 29\\
70569 Stuttgart, Germany}
\affiliation{SOFIA Science Center\\
NASA Ames Research Center\\
Mail Stop N211-1\\
Moffett Field, CA 94035, USA}

\author{Mark Lewis}
\affiliation{Trinity University\\
Dept. of Computer Science\\
One Trinity Place\\
San Antonio, TX 78212, USA}

\author{Karsten Schindler}
\affiliation{Deutsches SOFIA Institut\\
Universit\"at Stuttgart\\
Pfaffenwaldring 29\\
70569 Stuttgart, Germany}
\affiliation{SOFIA Science Center\\
NASA Ames Research Center\\
Mail Stop N211-1\\
Moffett Field, CA 94035, USA}



\begin{abstract}

A stellar occultation of Gaia DR3 2646598228351156352 by the Centaur (2060) Chiron was observed from the South African Astronomical Observatory on 2018 November 28 UT. Here we present a positive detection of material surrounding Chiron from the 74-in telescope for this event. Additionally, a global atmosphere is ruled out at the tens of $\mu$bar level for several possible atmospheric compositions. There are multiple $3\sigma$ drops in the 74-in light curve: three during immersion and two during emersion. Occulting material is located between $242-270$ km from the center of the nucleus in the sky plane. Assuming the ring-plane orientation proposed for Chiron from the 2011 occultation, the flux drops are located at 352, 344, and 316~km (immersion), and 357, and 364~km (emersion) from the center, with normal optical depths of 0.26, 0.36, and 0.22 (immersion) and 0.26 and 0.18 (emersion), and equivalent widths between 0.7-1.3~km. This detection is similar to the previously proposed two-ring system and is located within the error bars of that ring-pole plane; however, the normal optical depths are less than half of the previous values, and three features are detected on immersion. These results suggest that the properties of the surrounding material have evolved between the 2011, 2018, and 2022 observations.
\end{abstract}



\section{Introduction} \label{sec:intro}

The Centaur (2060) Chiron was discovered in 1977 \citep{RN1153, RN3998}. Its orbital semi-major axis ranges from approximately 8.5 to 19 AU, with an eccentricity of 0.38 and an inclination of \(7^\circ\). Chiron's rotational period has been reported as 5.92~h \citep[low error, by][]{RN1375}, with a slightly shorter period of approximately 5.5~h reported from more recent data \citep{RN3624,RN3657}. Chiron has exhibited outbursting behavior and is co-designated as comet 95P/Chiron \citep[e.g.][]{RN1146,RN824,RN1151,RN3558,RN3559}. Uncorrelated with heliocentric distance, Chiron’s brightness varies over both hourly and decadal timescales \citep[e.g.][]{RN834,RN3557}. For rotational variations, the typical peak-to-valley magnitude changes have not exceeded $0.09 \, {\rm mag}$ \citep{RN3557}. Chiron is one of the largest Centaurs: the Herschel Space Telescope determined a diameter of \(218 \pm 20 \, {\rm km}\) \citep{RN3624}, while data from the Atacama Large Millimeter Array returned a spherical equivalent diameter of $210^{+10}_{-12}$ km and a preferred elliptical size with semi-axes of \(114 \times 98 \times 62 \, {\rm km}\) \citep{RN3728}.

Stellar occultations provide a method that can directly and accurately determine the sizes and shapes of small bodies in the outer Solar System, as well as detecting and measuring atmospheres or rings \citep[e.g.][]{RN3785,RN3996}. Successful observations of stellar occultations by Chiron in 1993 and 1994 were consistent with a nucleus size of $>180$~km in diameter \citep{RN1386,RN1470}. Occultation data from 2018 and 2019 were more recently used to derive a Jacobi ellipsoid assuming a hydrostatic equilibrium shape, with semi-axes $a=126 \pm 22$~km, $b=109 \pm 19$~km, and $c=68 \pm 12$~km, implying a volume equivalent radius of \(R_{\rm vol} = 98 \pm 17 \, {\rm km}\) \citep{RN3992}.

 Notably, based on stellar occultation data from 2011, Chiron is one of only a few small bodies in the Solar System known to possibly host a ring system \citep{RN3716,RN3627}. A multi-chord stellar occultation in 2013 by the Centaur (10199) Chariklo revealed two, thin rings \citep{RN3626} and sparked a new field of study. Chariklo is the largest Centaur at $\sim250 \,{\rm km}$ in diameter, and the rings are roughly 385-390 and 400-405 km from the center of the nucleus, with widths of roughly 7 and 3 km and normal optical depths of 0.31-0.46 and 0.04-0.14, respectively \citep[e.g.][]{RN3626,RN3913}. A single, 70-km wide ring was later detected around the Trans-Neptunian Object (TNO) Haumea \citep{RN3677} and distant, inhomogenous material was recently reported around the TNO (50000) Quaoar \citep{RN3983,RN3989}. Haumea and Quaoar are an order of magnitude larger in size and located farther away from the Sun than Chariklo and Chiron. Nonetheless, these bodies share the traits of being vastly different in scales and compositions when compared to the giants previously known as the sole ring-bearing planets in our Solar System.  Study of each of these small bodies is thus merited. 

 In addition to providing an upper limit on Chiron's size, the stellar occultations in the 1990s detected what seemed jet-like features and an asymmetric dust coma around Chiron’s nucleus \citep{RN1386,RN1470}. The 2011 stellar occultation showed discrete, symmetric features on either side of the nucleus, which were initially interpreted as a near-circular arc or dust shell based on the limited, two-chord dataset \citep{RN3480}. \citet{RN3627} combined the 2011 occultation results with rotational light curves and long-term photometric and spectroscopic variations to propose that Chiron has a two-ring system similar to Chariklo, with mean radius of $324\pm10$ km and preferred pole orientation in ecliptic coordinates of $\lambda=144\pm10^\circ, \beta=24\pm10^\circ$. \citet{RN3716} further analyzed the 2011 occultation data, characterizing the proposed ring material to be at 300 and 309 km from the nucleus center in the ring plane and 2.5 to 4.5 km in width, given an assumed central chord. The ring widths varied azimuthally by 1.5 km (inner ring) and 0.5 km (outer ring) and the normal optical depths ranged between 0.6–0.85 (inner) and 0.5–0.71 (outer) \citep{RN3716}. Recently, \citet{RN3992} reported stellar occultations from 2018 and 2019 that did not show any features other than the nucleus: they concluded that a permanent ring similar in optical depth and extension to that at Chariklo was ruled out.

Here we report on data from the same 2018 stellar occultation by Chiron and at the same successful site of the South African Astronomical Observatory (SAAO) as discussed in \citet{RN3992}, but from a larger telescope. Surrounding material is detected in these data. The observations and data analyses are described in Sections \ref{sec:obs} and \ref{sec:analysis}. Results are presented in Section \ref{sec:results} and a discussion is provided in Section \ref{sec:disc}.

\section{Observations} \label{sec:obs}
\begin{figure}[ht!]
\centering
\includegraphics[width=0.3\textwidth,clip,trim=0mm 0mm 0mm 0mm]{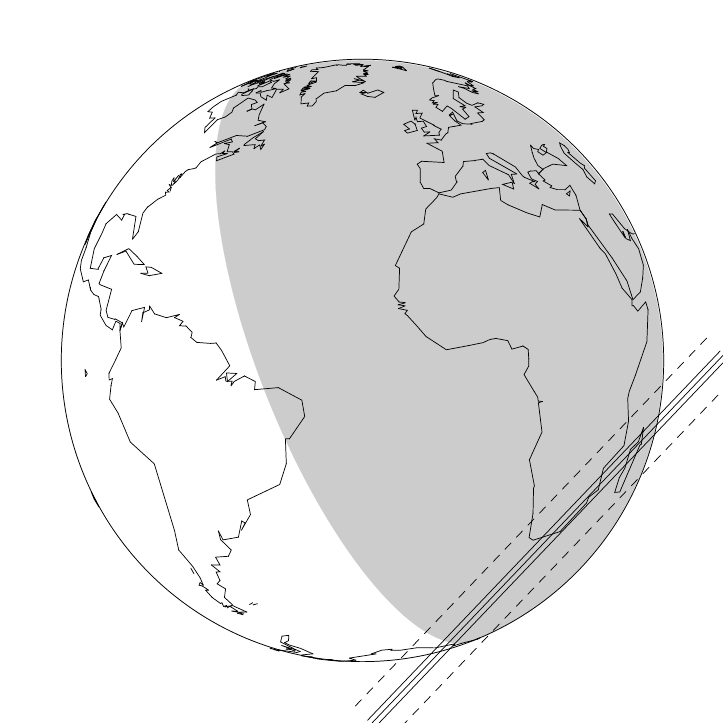}
\caption{The predicted shadow path for the 2018 November 28 UT Chiron occultation. The solid black lines indicate the northern, center, and southern extent of the shadow path with $3\sigma$ error bars as dashed lines. Chiron's shadow was assumed to be 108~km in radius for this prediction. 
\label{fig:globe}}
\end{figure}

Characteristics of the 2018 November 28 UT Chiron occultation star are listed in Table \ref{tab:star}. The prediction was based on the Gaia DR2 catalog, which was the most current star information at the time \citep{RN3687}, and Chiron's position was from the JPL Horizons Ephemeris System JPL$\# 121$ with offsets applied from our Ephemeris Correction Model \citep[ECM;][]{RN3646,RN3641,RN3632}. The predicted shadow path is shown in Fig. \ref{fig:globe}. This occultation was also predicted by the Lucky-Star project, using their Numerical Integration of the Motion of an Asteroid (NIMA), as described in \citet{RN3992} and \citet{RN3997}.

This occultation was predicted separately by two different groups. To allow independent campaigns and analyses, observations were planned on two telescopes at the same site: the 40 in and 74 in at the SAAO in Sutherland, South Africa (N latitude -32 22 54, E longitude 20 48 36, altitude 1760 m, WGS84, MPC site code K94). Two identical instruments were used on these telescopes, the Sutherland High-speed Optical Cameras \citep[SHOC,][]{RN3585}. The foundation of each system is an Andor iXon 888 camera system based around an e2v frame-transfer electron-multiplying charge-coupled device (EMCCD). Every occultation observation frame was triggered by a temperature-stabilized and frequency-tuned GPS. The timing accuracy is on the order of microsec. 

Data from the 40-in telescope taken with a cycle time of 1.5~s were reported in \citet{RN3992}. This work is focused on the 74-in dataset, although the 40-in data are also included here in order to show the light curve and to investigate light-curve stability. For the 74-in observations, the camera was cooled to $-50^\circ$C, no filters were used, and images were taken using the Conventional amplifier, read out at 1 MHz, and digitized with the 16-bit analog-to-digital converter. Data were taken with a cycle time of 0.5 s (including deadtime of 3.4 msec) from 20:20:00 to 21:20:00 UT. The field of view for SHOC on the 74-in telescope is roughly $1\farcm3 \times 1\farcm3$. Seeing varied throughout the observations from 2-3~arcsec. The images were binned $16 \times 16$ pixels for a plate scale of 1.22~arcsec.  
\begin{deluxetable*}{llcccccc}
\tablenum{1}
\tablecaption{Occultation star information.\label{tab:star}}
\tablewidth{0pt}
\tablehead{}
\startdata
Designation (GDR3)$^a$& 2646598228351156352 \\  
$\alpha$ (ICRS, Epoch Date; hh:mm:ss.ss)$^a$& 23:46:04.34\\
$\delta$ (ICRS, Epoch Date; ${}^\circ$:${}^\prime$:${}^{\prime\prime}$)$^a$ & 02:13:05.5\\
$\mu_\alpha$ (mas/yr)$^a$ & $-13.1 \pm 0.2$ \\
$\mu_\delta$ (mas/yr)$^a$ & $ -2.7 \pm 0.1 $ \\
Parallax (mas)$^a$ & $1.691 \pm 0.115$ \\ 
Magnitudes$^{a,b}$ & $G^{c}=17.3, G_{BP}=18.1, G_{RP}=16.3, J=15.2, H=14.8, K=14.3$ \\
Predicted midtime (UT)$^d$ & 2018 November 28, 20:50:59$\pm$00:01:07 \\
Predicted closest approach (km)$^d$ & $265\pm395$ N \\
Relative velocity$^e$ & 5.96 km/s\\
Position angle & $133\fdg7$ \\
\enddata
\tablecomments{$^a$From Gaia Data Release 3 \citep{RN3916}. $^b$Near infrared magnitudes from 2MASS \citep{RN3794,RN3995}. $^c$For comparison, the stars for the 2011 and 2022 Chiron occultations were $G=14.9$ and $G=12.7$, respectively. $^d$Based on Gaia Data Release 2 \citep{RN3687}, with $3\sigma$ error bars. $^e$Between the star and Chiron, from the SAAO.}
\end{deluxetable*}
\section{Data Analysis} \label{sec:analysis}

\begin{figure}[ht!]
\centering
\includegraphics[width=0.5\textwidth,clip,trim=0mm 0mm 0mm 0mm]{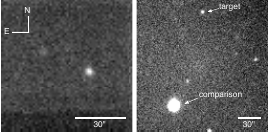}
\caption{Example images from the 74- and 40-in telescope data cubes. (left) The 74-in data, taken at 2-Hz cadence. The occultation star is the bright source, and a single, faint comparison ($G=18.7$) is detectable just above the noise to the NE. (right) The 40-in data, taken at 0.67-Hz cadence. The occultation star and a bright comparison star are labeled.
\label{fig:image}}
\end{figure}

\begin{figure}[ht!]
\centering
\includegraphics[width=0.4\textwidth,clip,trim=0mm 18mm 0mm 28.5mm]{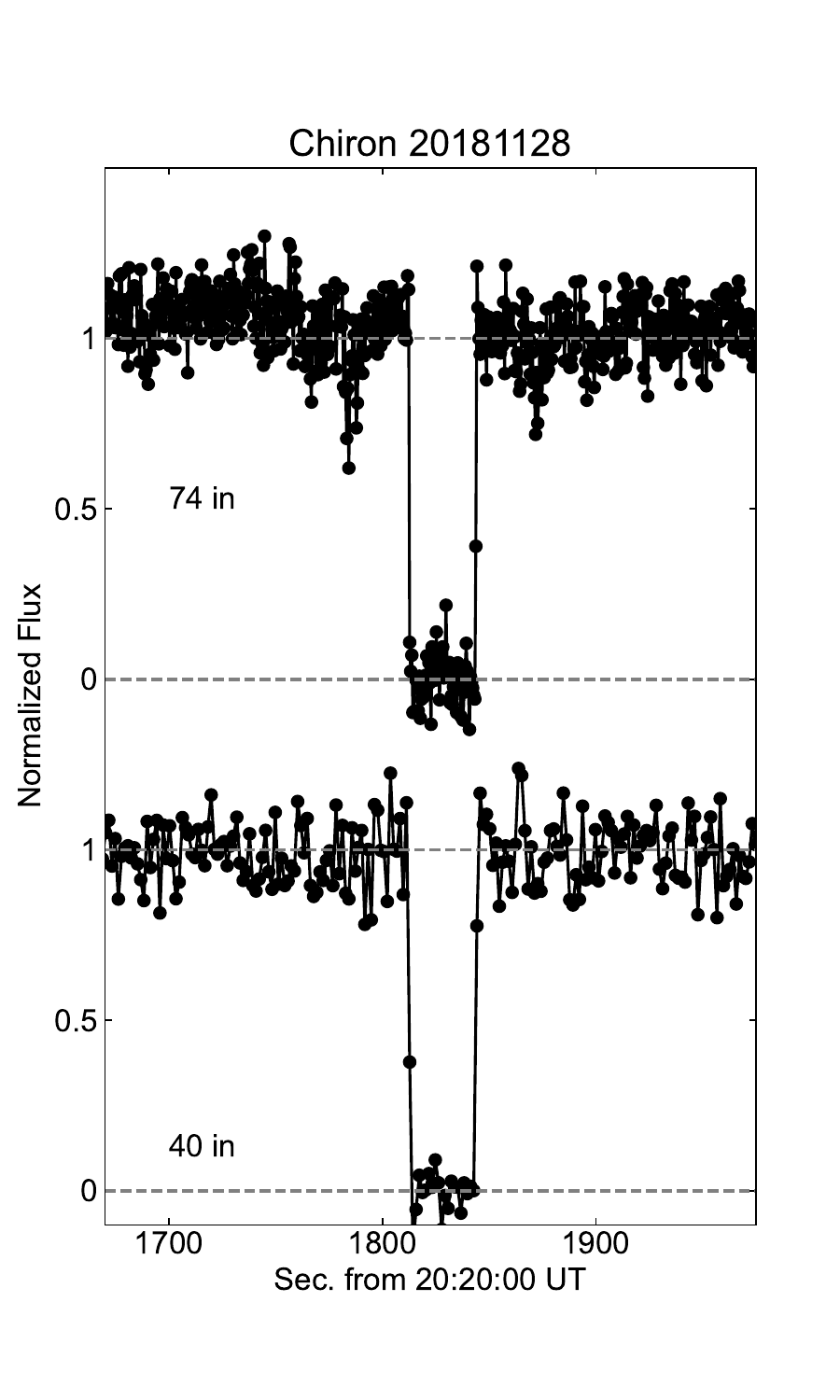}
\caption{Light curves from the (top) 74-in and (bottom) 40-in SAAO telescopes. The portion near the occultation by Chiron's nucleus is shown in more detail in Figs.~\ref{fig:3sigfeatures} and ~\ref{fig:atmfit}, with significant flux drops highlighted in Fig.~\ref{fig:ringfits}. 
\label{fig:lcs}}
\end{figure}

Raw data were analyzed in order to extract light curves from both the 40- and 74-in data, using circular-aperture photometry. Example images are shown in Fig. \ref{fig:image}. Four hours of time was requested on each telescope, centered on the occultation midtime; therefore, twilight time was not available to take flatfields. Reduced data using bias-subtraction alone or dithered sky flats added noise, in terms of increasing the standard deviations of the light curves. Dark images were not required as the camera cooling has undetectable dark current at these exposure times \citep{RN3585}. For the 40-in analysis, the comparison star was centroided on each frame and an offset was applied to determine the location of the target star. Photometry was carried out on both sources. For the 74-in analysis, the field of view was too small to contain a comparison star of adequate magnitude (see Section \ref{sec:disc} for additional discussion). Photometry was carried out on the target star alone, which was centroided on each image. For both datasets, four, hand-selected boxes well outside of the apertures and without background stars were averaged to determine the sky value, which was subtracted out. A sequence of aperture sizes were tested, and the highest signal-to-noise ratios (SNRs) were obtained with apertures of 8 superpixels (9.8 arcsec) and 6 superpixels (8.0 arcsec) in diameter for the 74-in and 40-in respectively. For the 40", the target star signal was divided by the comparison star for differential photometry. The light curves were normalized to one by dividing by the mean value of the baseline (flux=1) and scaled such that the occulted portion had a mean value of zero. Data outside of the occultation were fit by a third-order polynomial to ensure a flat baseline. The resulting light curves are shown in Fig.~\ref{fig:lcs}. There is an obvious detection of Chiron's nucleus in both datasets, and there are significant dips in the 74-in light curve that are not present in the 40-in data. 

The important size scales for diffraction effects are the Fresnel length, $\sqrt{\lambda D/2}$, the stellar diameter, and the integration length of the observations (integration time multiplied by the velocity in the sky plane). $\lambda$ is the wavelength of the observation, here assumed to be $700\, {\rm nm}$, and $D$ is the distance from Earth, 18.355 AU at the midtime of the occultation. For this event, the Fresnel length is $0.98 \, {\rm km}$. The maximum stellar diameter is $0.1 \, {\rm km}$: the same value is obtained using the TESS stellar diameter \citep{RN3976} projected to Chiron's distance or using NOMAD magnitudes \citep{RN3408} and the equations for the angular size of a giant star from \citet{RN2265}. The lengths of the integrations correspond to approximately $3 \, {\rm km}$ and $9 \, {\rm km}$ for the 74-in and 40-in data, respectively. The diffraction effects are thus dominated by the integration times and the stellar profile can be neglected.

We fit the immersion and emersion of the star behind the nucleus following \citet{RN89}, for a thin atmosphere with diffraction on a solid surface. This model is based on a radially-symmetric atmosphere with constant scale height much less than the body's radius. The model is applicable when the scale height is much greater than the Fresnel scale, which is the case for this event. Immersion and emersion are fit separately. The fits are carried out at three representative scale heights, 10, 50, and 100 km. The fit parameters are the zero and full light-curve levels, the geometric edge, and the bending parameter, $b$, which represents the amount of differential bending of light causing a change in flux at the surface. The maximum $3\sigma$ bending parameter over all of the tested scale heights is used as the upper limit on the flux change at the surface of Chiron. We then consider the maximum bending parameter along with different constituent molecules to place corresponding upper limits on a possible atmosphere. For the atmospheric calculations, we assume a temperature of $125 \, {\rm K}$ \citep{RN1429} and a mass of $4.8 \pm 2.3 \times10^{18}\, {\rm kg}$ and volume equivalent radius of $98 \pm 17 \, {\rm km}$ from \citet{RN3992}. We use this volume equivalent radius in order to have the best estimate of Chiron's gravitational pull. The errors on the mass and size of Chiron's nucleus are carried through when calculating the atmospheric upper limits and their errors.

\begin{figure}[ht!]
\centering
\includegraphics[width=0.4\textwidth,clip,trim=0mm 0mm 0mm 0mm]{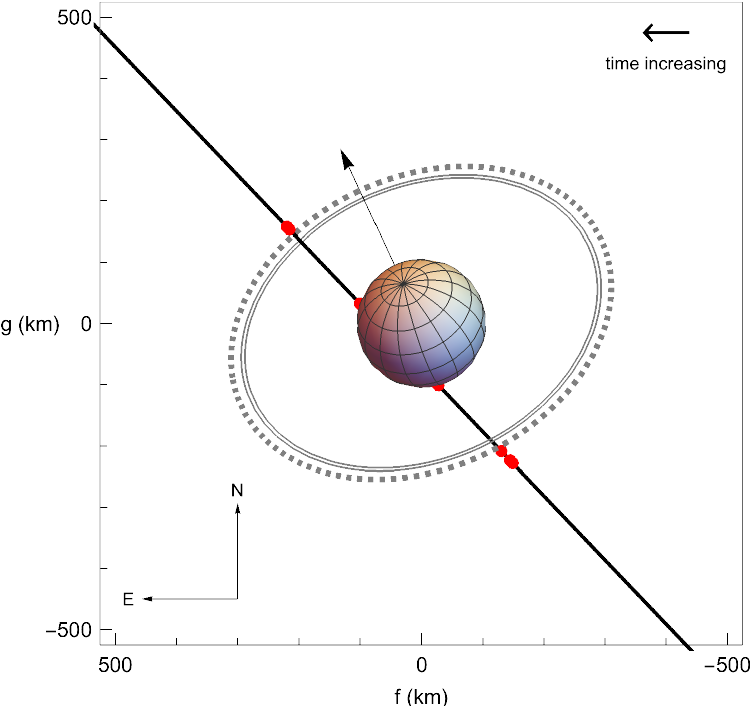}
\caption{Sky-plane projection of Chiron at the time of the occultation, along with the nominal location of the observed chord and the proposed rings. Chiron’s orientation is that of the preferred ring-plane pole from \citet{RN3627} and the chord has a closest approach of 50 km calculated from assuming a spherical nucleus with a radius of 105 km (following \citet{RN3728}) and using the ingress and egress times in Table \ref{tab:difffitd}. Each black dot represents one image taken using the 74-in telescope, which appear continuous at this resolution. Enlarged, red dots indicate locations where the light-curve data deviate by $3\sigma$ or more below the baseline, with the chord going behind the solid nucleus in this plot. Solid gray ellipses represent the two most prominent dips from \citep{RN3716}, while the dotted gray ellipse represents the ring location and width from \citet{RN3627}.}  
\label{fig:skyplane}
\end{figure}

The closest-approach of the chord to the center of the nucleus was determined to be 50 km by assuming a sphere of 210 km in diameter \citep{RN3728} and using the measured ingress and egress times from the diffraction fitting (results shown in Table \ref{tab:difffitd}). This geometry is consistent with that reported in \citet[][; their Fig. 9]{RN3992}, in which a sphere is shown in the sky-plane figure. Because there is only a single chord for this event, a unique solution for the closest approach by a body of unknown shape cannot be determined. Chiron is likely to be a triaxial ellipsoid, values for which were constrained with large error bars by \citet{RN3992}. Assuming a spherical nucleus is a good first assumption, for example following \citet{RN4001} and including error bars on the closest approach to allow for a different nucleus shape. See Section \ref{sec:disc} for additional discussion. The sky-plane view for the nominal occultation geometry is shown in Fig.~\ref{fig:skyplane}. As noted in \citet{RN3992}, the null detection from Boyden Observatory for this event dictates that the chord passed south of the center.

\begin{figure}[ht!]
\centering
\includegraphics[width=0.5\textwidth,clip,trim=0mm 0mm 0mm 0mm]{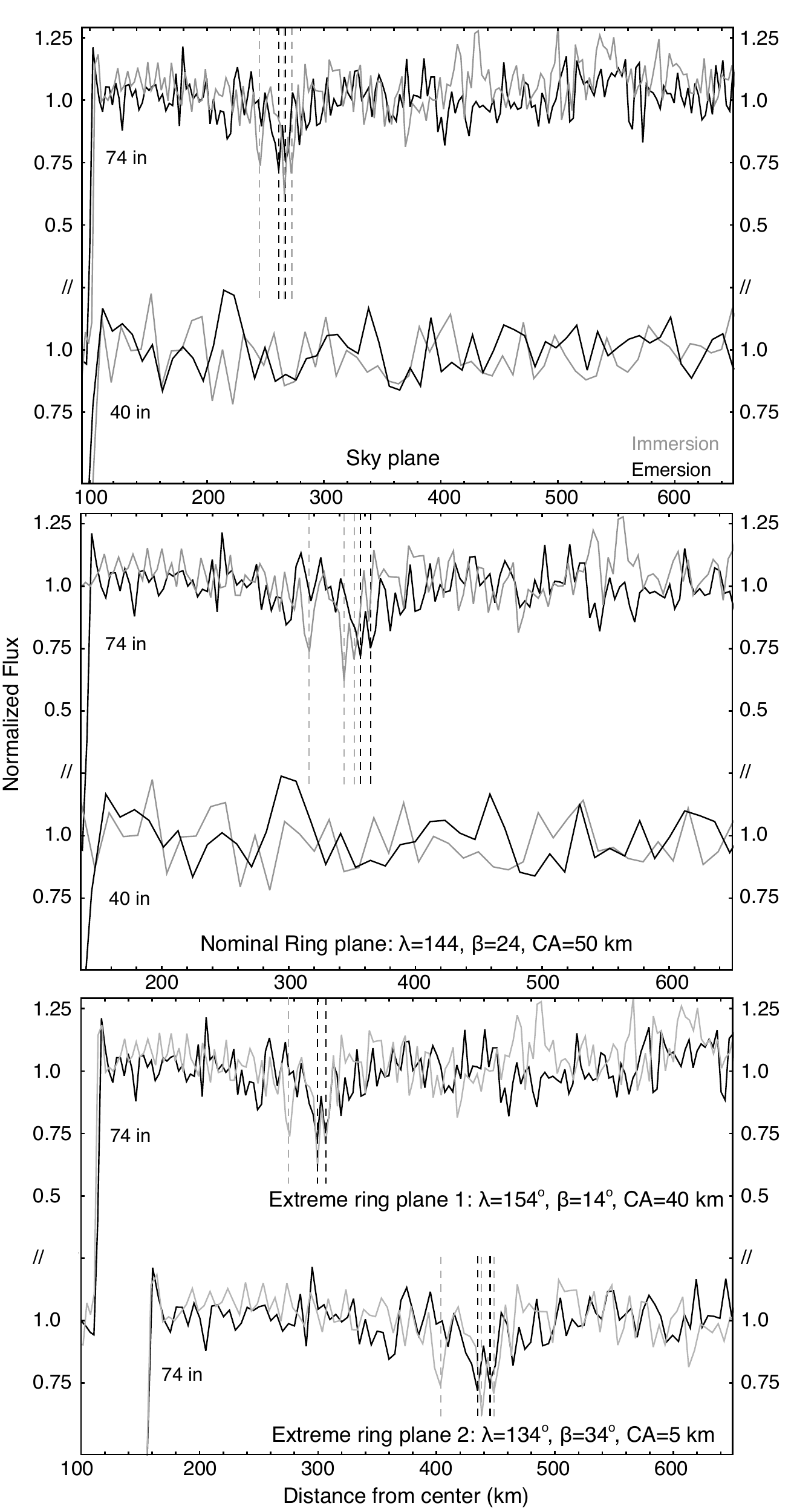}
\caption{Test for symmetric features. The immersion (gray) and emersion (black) portions of the light curves are overlaid to check for features that are symmetric on either side of the nucleus. The overlay is centered on the image closest to the fitted midtime. Dashed lines are provided at the locations of all data points that are $3\sigma$ or more below the baseline. (top) Data with respect to distance in the sky plane. (middle) Data with respect to distance in the nominal ring plane, with the preferred ring-orbital pole from \citet{RN3627}. In both panels, data from the 74-in telescope are shown on the top and data from the 40-in telescope are shown on the bottom. No significant features were detected in the 40-in data. (bottom) Example alternative geometries for the 74-in dataset: data with respect to the ring plane for the extremes of the ring-orbital pole errors in \citet{RN3627}. In order for the nucleus immersion and emersion to align in this panel, the closest-approach (CA) value was adjusted from the nominal value to the values given in the plot.
\label{fig:3sigfeatures}}
\end{figure}

To search for symmetric features around Chiron in the light curve, we plot the immersion and emersion portions versus distance from the body center and overlay them in Fig.~\ref{fig:3sigfeatures}. The standard deviations of the baselines are 0.08 and 0.09 normalized flux for the 74-in and 40-in data, respectively. Features that are greater than or equal to three-sigma drops from the baseline are flagged. There are three such features during immersion and two during emersion (as highlighted in red in Fig. \ref{fig:skyplane}). The top and middle panels of Fig.~\ref{fig:3sigfeatures} respectively demonstrate the data in the sky-plane and with the nominal ring-plane geometry. As discussed in Section \ref{sec:disc}, the bottom panel explores a wider range of possible ring-pole positions and closest-approach values.

We fit each of the $3\sigma$ deviations in the sky plane using a square-well diffraction model for occultation profiles that treats the blocking material as a uniformly transmitting grey screen with sharp edges, following \citet{RN391}. Fit parameters for the model are the apparent optical depth, $\tau$, and the width in the sky plane. Distances from the nucleus and widths in the nominal ring-plane are determined using the preferred solution of the proposed Chiron ring system from \citet{RN3627} and our best-fit closest approach of 50 km. We convert to normal optical depths, $\tau_{\rm N}$, and equivalent widths by using the opening angle from the preferred ring solution, $B = 47{\fdg}3$.

\section{Results} \label{sec:results}

\begin{figure}[ht!]
\centering
\includegraphics[width=0.7\textwidth,clip,trim=0mm 0mm 0mm 0mm]{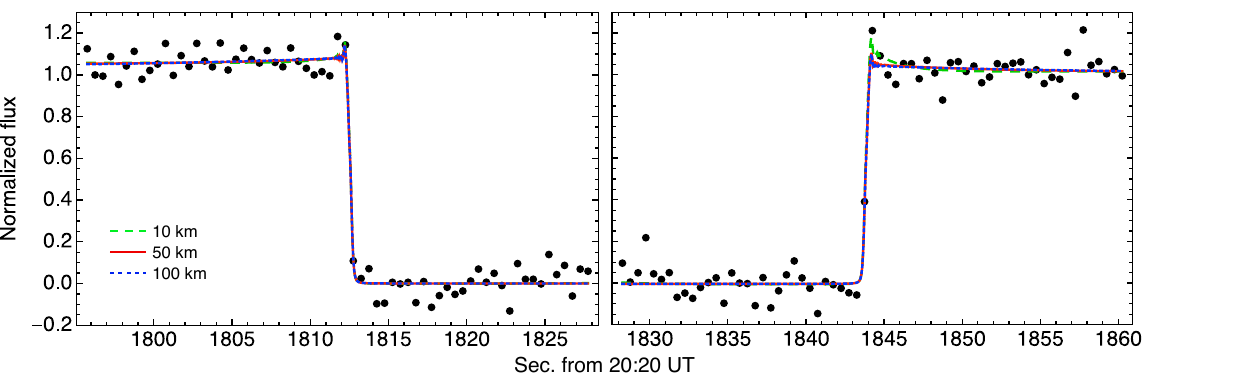}
\caption{Diffraction model fit to the 74-in data for immersion (left) and emersion (right). Data are shown as black dots. The best-fit models are shown as dashed green, solid red, and dotted blue lines for 10, 50, and 100-km scale heights, respectively. The models are very similar, with a slightly larger diffraction spike on emersion for the 10-km scale height (consistent with the lowest bending parameter, $b$, for this case in Table \ref{tab:difffitd}).
\label{fig:atmfit}}
\end{figure}

The results from the diffraction fitting of the nucleus are plotted in Fig. \ref{fig:atmfit} and listed in Table \ref{tab:difffitd}. The nucleus occultation lasted 31.23~s, or slightly over 186~km at Chiron. The maximum $3\sigma$ bending parameter is 0.13 on immersion and 0.17 on emersion (the maximum values of $b$ in Table \ref{tab:difffitd} considering $+3\sigma$). The considered atmospheric constituents and limits on an atmosphere at immersion and emersion are listed in Table \ref{tab:atmos}. These $3\sigma$ upper limits are roughly between 10 and 30 $\mu$bar, with constraints on emersion being approximately $20\%$ larger than those on immersion.
\begin{deluxetable*}{llcccccc}
\tablenum{2}
\tablecaption{Results from diffraction fitting to the nucleus.\label{tab:difffitd}}
\tablewidth{0pt}
\tablehead{}
\startdata
Immersion, geometric edge (UT)$^a$ & $ \hbox{20:50:12.58} \pm 0.04 \, {\rm s}$ \\
Emersion, geometric edge (UT)$^a$  & $ \hbox{20:50:43.81} \pm 0.02 \, {\rm s}$ \\
Occultation duration (s) & $31.23 \pm 0.04$ \\
Chord length (km) & $186.13 \pm 0.27$ \\
Immersion $b$ (10, 50, 100 km; $10^{-2}$) & $-4.3 \pm 5.6,-3.1 \pm 4.3,-4.7 \pm 5.9$ \\
Emersion $b$ (10, 50, 100 km; $10^{-2}$) & $-12.6 \pm 6.3,-3.8 \pm 5.1,-4.0 \pm 6.9$ \\
\enddata
\tablecomments{$^a$Following \citet{RN89}.}
\end{deluxetable*}
\begin{deluxetable*}{llcccccc}
\tablenum{3}
\tablecaption{$3\sigma$ upper limits on a possible Chiron atmosphere.\label{tab:atmos}}
\tablewidth{0pt}
\tablehead{
\colhead{Molecule} & \colhead{Immersion ($\mu$bar)} &\colhead{Emersion ($\mu$bar)}
}
\startdata
CH\textsubscript{4}  & 12.7& 16.5 \\
CO                  & 22.1& 28.7 \\
CO\textsubscript{2} & 20.7& 26.9 \\
H\textsubscript{2}  & 14.3& 18.6 \\
H\textsubscript{2}O  & 23.6& 30.7 \\
N\textsubscript{2}  & 24.9& 32.4 \\
\enddata
\end{deluxetable*}

\begin{figure}[ht!]
\centering
\includegraphics[width=0.7\textwidth,clip,trim=0mm 0mm 0mm 0mm]{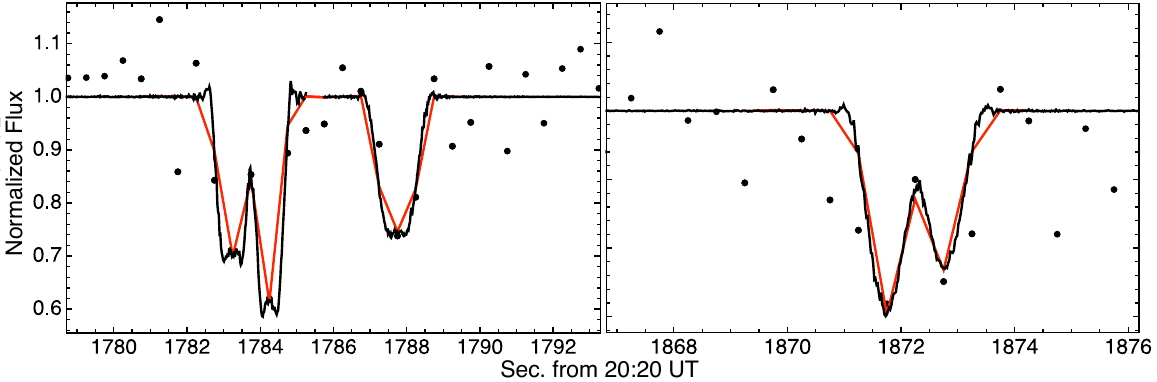}
\caption{Diffraction model fits to the $3\sigma$ drops in the 74-in light curve. Dots represent data points and lines represent the best-fit models. The black lines are the diffraction models at high-time resolution, and red lines represent the model fits at the sampling rate of the dataset. Overlapping features were fit simultaneously.
\label{fig:ringfits}}
\end{figure}

Figure \ref{fig:ringfits} shows the portions of the light curve containing the $3\sigma$ dips in the light curve, along with plots of the best-fit diffraction models. Table~\ref{tab:features} contains the measured and fitted characteristics for each of these significant features. The double-dips, two overlapping features on both immersion and emersion, are located between 344-365 km from the center of the nucleus in the nominal ring plane. There is an approximately 8~km gap between the centers of the two features on each side, although the immersion dips are roughly 13~km closer to the nucleus than the emersion. The immersion dips are wider, at 4.5- and 4.0-km for the exterior and interior dips, respectively, versus 2.2-km ring-plane widths for both on emersion. The normal optical depths are also slightly higher on immersion, at 0.26 and 0.36 versus 0.19 and 0.26. The third, stand-alone dip on immersion is closer to the nucleus, at 317~km from the center. It is the widest feature, at 5.1~km and has a similar normal optical depth to the other dips of 0.22.
\begin{deluxetable*}{cccccccc}
\tablenum{4}
\tablecaption{Characteristics of significant light-curve features.\label{tab:features}}
\tablewidth{0pt}
\tablehead{
\colhead{} & \colhead{} & \colhead{Sky-plane} & \colhead{Line-of-sight}& \colhead{Ring-plane}& \colhead{Ring-plane}& \colhead{Normal}& \colhead{Equivalent}\\
\colhead{Midtime} & \colhead{Distance} & \colhead{width} & \colhead{optical}& \colhead{distance}& \colhead{width}& \colhead{optical depth}& \colhead{width}\\
\colhead{(sec after 20:20 UT)} & \colhead{(km)${}^a$} & \colhead{(km)${}^b$} & \colhead{depth,$\tau$${}^b$}& \colhead{(km)${}^c$}& \colhead{(km)${}^c$}& \colhead{$\tau_N$${}^d$}& \colhead{(km)${}^e$}
}
\startdata
1783.25 & 269.16 &  $5.19\pm{0.26}$	& $0.36\pm{0.11}$	&351.50& 	$4.54\pm{0.23}$ & $0.26\pm{0.08}$	&	$1.16\pm{0.36}$ \\
1784.25 & 263.19 &  $4.59\pm{0.30}$	& $0.46\pm{0.11}$	&343.67&	$4.03\pm{0.26}$	&  $0.36\pm{0.08}$	&	$1.24\pm{0.31}$ \\
1787.75 & 242.33 &  $5.76\pm{0.57}$	& $0.30\pm{0.10}$	&316.30&	$5.10\pm{0.51}$ &   $0.22\pm{0.07}$	&	$1.11\pm{0.42}$ \\
1871.75 & 258.31 &  $3.37\pm{0.75}$	& $0.35\pm{0.10}$	&356.64&	$2.23\pm{0.50}$	&	$0.26\pm{0.07}$   &$0.76\pm{0.33}$ \\
1872.75 & 264.27 &  $3.36\pm{0.78}$	& $0.25\pm{0.14}$	&364.48&	$2.23\pm{0.52}$	&	$0.19\pm{0.07}$   &$0.67\pm{0.32}$ \\
\enddata
\tablecomments{${}^a$Distance from the center of the nucleus, as measured in the sky plane. Error is $\pm$ half of a time step: 0.25 s or 1.49 km. ${}^b$Based on least-squares fits to the data using a square-well diffraction model for occultation profiles from \citet{RN391}. Error bars were determined by fitting the light-curve data at $\pm1\sigma$ flux. ${}^c$The conversion from the sky plane to the ring plane assumes the preferred pole position from \citet{RN3627} with a closest approach of 50 km. ${}^d$Normal optical depths assume a ring opening angle of $B = 47{\fdg}3$, where $\tau_{\rm N}=\tau \sin|B|$. ${}^e$Calculated by multiplying the width in the sky plane by $(1-e^{-\tau})\sin|B|$.}
\end{deluxetable*}
\section{Discussion} \label{sec:disc}
Observations of this 2018 November 28 UT stellar occultation by Chiron were successful from two telescopes at the SAAO. An occultation by the nucleus was observed from each telescope. No other significant flux drops were detected in the 40-in dataset \citep[as reported in][and shown in Fig. \ref{fig:3sigfeatures}]{RN3992}, while five significant flux drops are reported here in the 74-in dataset. This event was the first Chiron occultation observation since 2011 in which surrounding material has been detected.

From the 74-in data, the occultation by the nucleus has a chord length of $186.1\pm{0.3} \, {\rm km}$. This size is a few km longer than the value obtained in \citet{RN3992}; however, (i) we use a diffraction fit in which the geometric edge is located near 0.25 flux in the shadow (derived from the diffraction model for a knife edge) and (ii) based on Table 5 in \citet{RN3992}, a different relative velocity was assumed than the 5.96~km/s that we have calculated at the SAAO (the egress-ingress times and chord length in their Table 5 imply a velocity of 5.77 km/s). Our chord length is consistent with previous size estimates for Chiron.

\begin{figure}[ht!]
\centering
\includegraphics[width=0.6\textwidth,clip,trim=0mm 0mm 0mm 0mm]{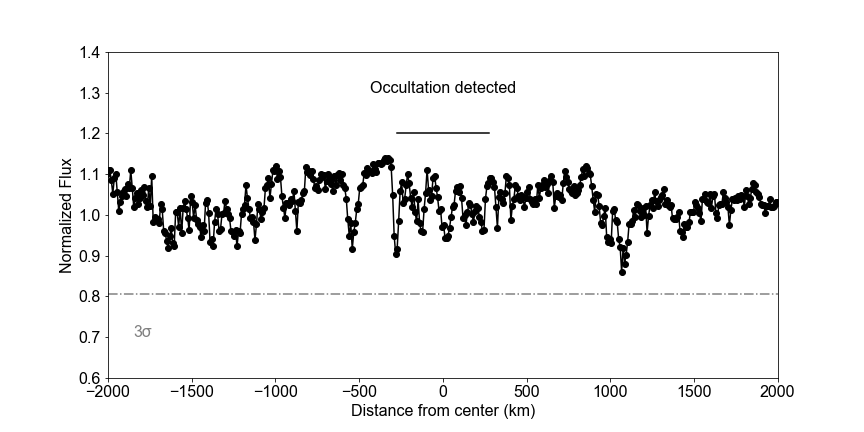}
\caption{Light curve for the comparison star in the 40-in dataset. The dot-dashed line indicates the level of $3\sigma$ drops from the baseline. The solid line shows the spatial extent of the occultation features discussed in this work.
\label{fig:complc}}
\end{figure}

The 74-in field of view was too small to provide an adequate comparison star for differential photometry. Therefore, care must be taken to prevent misinterpreting variations in flux that are due to changes in sky conditions as opposed to occultations at the targeted body. The detected flux drops are located near the proposed ring locations, and we find no other $3\sigma$ drops in the light curve out to more than a thousand km from the nucleus. In addition, the 40-in dataset did have a bright comparison star, and there were no $3\sigma$ flux drops in that star's signal within a few thousand km on either side of the occultation midtime (see Fig. \ref{fig:complc}). We thus determine that the star signal was sufficiently stable during the occultation that the detected drops were caused by material near Chiron.

\begin{figure}[ht!]
\centering
\includegraphics[width=0.8\textwidth,clip,trim=0mm 0mm 0mm 0mm]{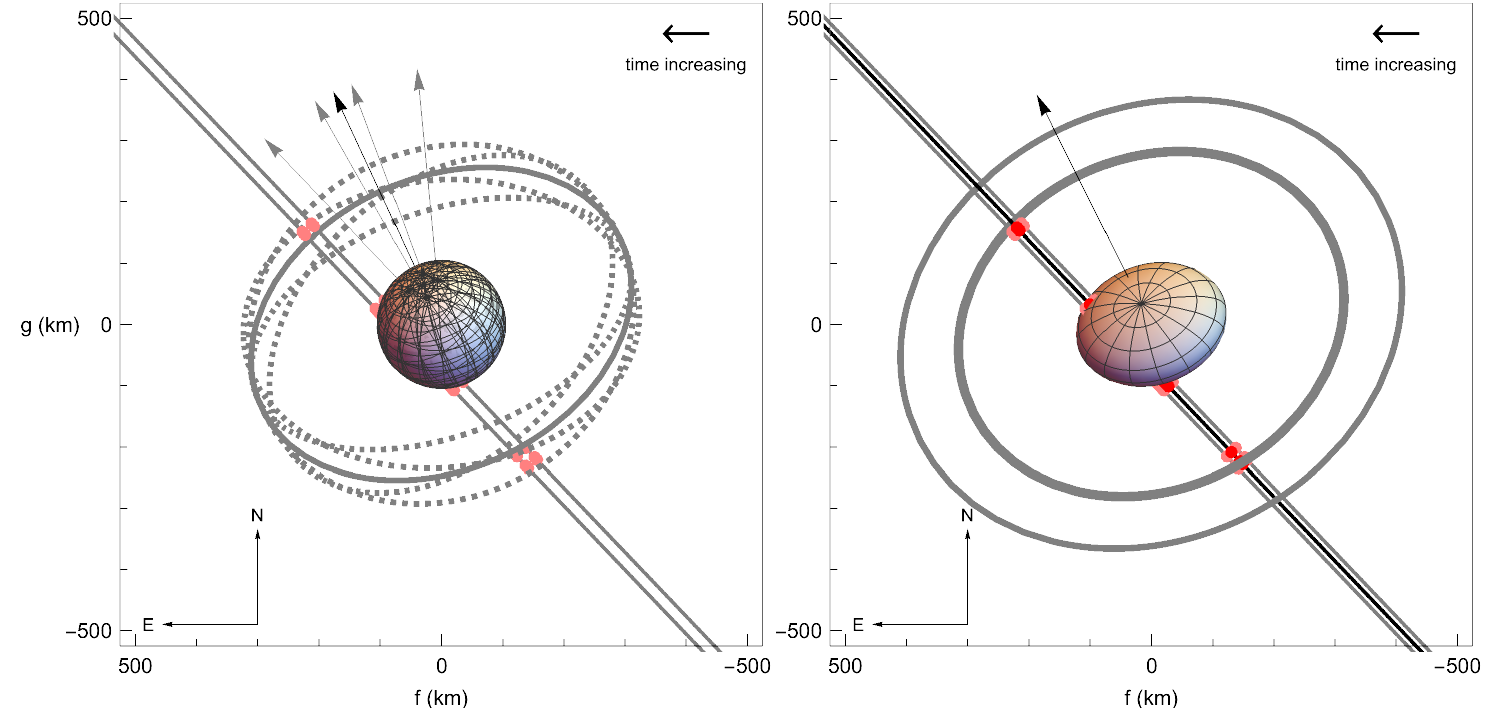}
\caption{Sky-plane projections of Chiron at the time of the 2018 occultation, along with the location of the observed chord for different ring-pole positions and closest approaches. Each black or gray dot represents one image taken using the 74-in telescope, which appear continuous at this resolution. Enlarged, red or pink dots indicate locations where the light-curve data deviate by $3\sigma$ or more below the baseline, with the chord going behind the solid nucleus in these plots. (left) Range of ring-pole positions based on the errors on ecliptic latitude and longitude for the preferred solution in \citet{RN3627}. The solid gray ellipse represents the ring location for the preferred pole position (the pole indicated by a black arrow, as shown in Fig.~\ref{fig:skyplane}). Dotted gray ellipses indicate the different ring locations for the extremes of the ring-pole error bars (the poles indicated by gray arrows). The chords are plotted with closest approach distances $\pm10 \, {\rm km}$ from the nominal 50-km solution, each data point is plotted in gray with the significant light-curve deviations represented in pink. (right) The best-fit triaxial shape for Chiron's nucleus from \citet{RN3992} with the updated nominal ring-pole position (not including  errors) and ring concentrations (including errors) from \citet{RN4001}. The chord is plotted in black and red for the nominal closest approach with gray and pink for closest approaches of $\pm10 \, {\rm km}$.}  
\label{fig:skyplane2}
\end{figure}

Fig.~\ref{fig:skyplane} shows the nominal occultation geometry for this event. In the nominal ring plane, the 2018 double-dip features are outside of the previously-proposed distances from the nucleus, at 20-40 km exterior to the $324 \pm{10} \, {\rm km}$ ring location from \citet{RN3627}. However, the geometry varies depending on the exact location of the ring-pole position and the closest-approach distance of the chord. \citet{RN4001} recently presented results from a stellar occultation by Chiron in 2022 December. They found that the structure of the surrounding material evolved since 2011. The 2022 data indicated a broad, tenuous disk of 580 km with ecliptic coordinates of the pole being $\lambda=151^\circ \pm{8^\circ}$ and $\beta=18^\circ \pm{11^\circ}$ as well as concentrations of material at $325\pm{16} \,{\rm km}$ and $423\pm{11} \,{\rm km}$ \citep{RN4001} . This pole position is within the $10^{\circ}$ errors of the previous estimate and has similarly-sized error bars. To compare occultation geometries between 2011 and 2022, \citet{RN4001} assumed a spherical nucleus of the same size assumed in this work with a radial uncertainty of 10 km to account for any displacement due to a triaxial shape. Sky-plane projections are given in Fig.~\ref{fig:skyplane2} to explore the impact of different, possible ring geometries for the 2018 dataset. The left panel in Fig.~\ref{fig:skyplane2} shows the range of ring and ring-pole positions including the error bars from the preferred solution in \citet{RN3627} along with the 2018 chord closest approach shifted by $\pm10 \, {\rm km}$. The right panel in Fig.~\ref{fig:skyplane2} shows the triaxial nucleus shape from \citet{RN3992} with the newest ring-pole position and the locations of ring concentrations from \citet{RN4001}. Changing the closest approach (the equivalent of considering a non-spherical nucleus) moves the locations of the significant features by a maximum of 1~km on ingress and 7~km on egress for the nominal ring plane. The possible locations in the sky-plane for ring material are much broader when considering the ring-pole position errors (left panel in Fig.~\ref{fig:skyplane2}). 

Likewise, the bottom panel in Fig. \ref{fig:3sigfeatures} provides an example of how the locations of significant features can vary as a function of the ring-pole position: two extreme cases at the edges of the error bars of the ring-pole-position ecliptic coordinates are plotted in the bottom panel. These each require lower closest-approach distances than the nominal value in order to align the occultation by the nucleus. The locations of the significant features in the ring planes change by more than 100 km, depending on the ring-pole location. For extreme case 1, the 2018 dip-double features appear well aligned at approximately 300 and 306 km from the center. This orientation falls within the error bars of the newer \citet{RN4001} ring-pole position. For extreme case 2, the nucleus would be more than 300 km in diameter in the ring-plane, which is not consistent with any reported size measurements for Chiron. While we can test for locations of surrounding material given a nominal ring-pole position, more accurate ring-pole coordinates are needed to narrow down the range of possible locations and more accurately measure positional changes over time.

The occultation star in 2018 was significantly fainter than in 2011 or 2022 (see Table \ref{tab:star}). SNRs of occultation light curves are typically reported as the mean over the standard deviation of the normalized flux: to compare data quality between datasets, the spatial scale of the sampling must also be considered. For example, for the 2018 occultation data the 0.5-s cadence at the 74-in telescope had a spatial sampling of 2.98 km compared to the 1.5-s or 8.9-km sampling in the 40-in dataset. We select 3 km as a comparative spatial scale, as the highest resolution ring widths at Chiron from \citet{RN3716}. The 2018 datasets have a SNR per 3 km at Chiron of 12.5 and 7.2 for the 74-in and 40-in, respectively. The SNR was insufficient to detect material in the 2018 data from the 40-in telescope while the material was apparent in the 74-in dataset. In 2011, the velocity was $9.87 \,{\rm km s^{-1}}$ and the SNRs per 3 km were 25.9 and 18.7 for the 0.2-s FTN and 0.5-s MORIS data, respectively \citep{RN3716}. In 2022, the velocity was $4.25 \,{\rm km s^{-1}}$ and the highest-quality dataset was from Kottamia Observatory with a sampling of 4.5 s and a standard deviation of 0.01 estimated from Fig. 1 of \citet{RN4001}, corresponding to a robust SNR per 3 km of 40.

From the 2011 data, two, distinct drops in flux with ring-plane widths of $2.6-4.4 \, {\rm km}$ and normal optical depths of $0.5-0.85$ were reported \citep{RN3480,RN3716}. The highest-quality 2011 light curve had a spatial resolution of 2 km at Chiron. The best 2018 light curve has spatial resolution of 3 km at Chiron, and the significant drops have similar widths but at lower optical depths than those detected in 2011. The 2022 data have a resolution of only 19 km: optical depths were not reported, but the depth and breadth of flux attenuation \citep{RN4001} suggests higher optical depths in 2022 for features that would have similar widths to those observed in 2018. Indeed, the detection of surrounding material from 2022 led to the conclusion by \citet{RN4001} that there was more material present than in 2011. In terms of location, the 2018 double-dip features are nominally outside of the previously-proposed distance from the nucleus (see Fig. \ref{fig:skyplane}); however, they do fall within the range of possible locations given the error bars on the ring-pole position (see the left panel of Fig. \ref{fig:skyplane2}). The 2018 features are slightly interior to the newly-determined ring concentrations from \citet{RN4001} (see right panel of Fig. \ref{fig:skyplane2}), noting that the errors in the newer ring-pole position are similar to those from 2011 and thus likely encompass the 2018 detections. The gap between the two features remained roughly the same at 8 km in 2018 and 9-10 km in 2011 \citep{RN3716} \--- the resolution was too low to observe such a gap in the 2022 data \citet{RN4001}. An additional, single drop in flux was detected in 2018 within the proposed ring location at 316 km from the center in the nominal ring plane. \citet{RN3716} also reported one drop in flux on immersion that was interior to the proposed two-ring system, but it was broader (17.6 km ring-plane width) and more optically thick ($\tau_{\rm N}=0.29$) than the third feature on immersion observed here. 

The orbital distances and widths of the flux drops at Chiron are roughly similar to Chariklo's established rings, with the latter system at $\sim400 \, {\rm km}$ and widths of $\sim3$ and 7 ~km \citep{RN3626}. Note that observed values of optical depths for small-body rings have not been consistently calculated in the literature. Published values for optical depths for material surrounding Chiron \citep{RN3716} and in this work are based on the standard relationship of ${I}/{I_0}=e^{-\tau}$, where $I$ and $I_0$ are the transmitted and incident fluxes. The normal optical depth is calculated using the opening angle, $B$, by $\tau_{\rm N}=\tau \sin|B|$ for a polylayer ring \citep[e.g.][]{RN391}. Reported optical depths for Chariklo's rings, the proposed ring at Quaoar, and the recent recalculation of the characteristics of the features at Chiron have included a factor of two, $\tau_{\rm N}=\frac{\tau}{2} \sin|B|$ \citep{RN3626,RN3658,RN3983,RN3992}. According to \citet{RN3658}, the factor of two stems from \citet{RN2155}, who found that the optical thickness measured from stellar occultations for the Uranian rings was larger by an extinction efficiency factor of two than the fractional area physically filled by particles \citep[i.e. the Mie coefficient was $Q_{\rm e}=2$,][]{RN2155}. Typically, the Mie coefficient comes into play when inferring physical characteristics of ring particles, which we have not attempted here.  This factor of two was noted as the primary difference between optical depths reported in \citet{RN3716} and the recalculated versions in \citet{RN3992}. This difference also means that the $\tau_N$ values reported for Chariklo need to be increased by a factor of two for direct comparison with the reported values for Chiron from this work. Here, we find normal optical depths of roughly 0.2-0.4, and the comparable values for Chariklo's rings would be such that C1R is much more optically thick at $\tau_N=0.8$ and C2R is more optically thin at $\tau_N=0.12$. 

Ring-type material around small bodies in the outer Solar System has been discovered well beyond the Roche limit, a location beyond which basic theory would suggest that particles should disperse or accrete \citep[e.g.][]{RN3983,RN3989,RN3815}. The Roche radius can be defined as $a_{Roche}=(3/\gamma)^{1/3}(M/\rho^\prime)^{1/3}$, where $M$ is the body's mass, $\rho^\prime$ is the density of the orbiting material, and the factor describing particle shape $\gamma=0.85$ for classical calculations, while $\gamma=1.6$ has been preferred to represent the tidal destruction limit \citep[e.g.][]{RN3983,RN3712}. For Chiron's mass $M=4.8\pm{2.3}\times10^{18} \, {\rm kg}$ \citep{RN3992}, water ice particles with $\rho^\prime=1 \, {\rm g \, cm^{-3}}$, and either value of $\gamma$, Chiron's Roche limit is $<260$ km. Considering the extreme case of the largest mass including errors and $\rho^\prime=0.45 \, {\rm g \, cm^{-3}}$, the density of Saturnian satellites \citep{RN3984}, the Roche limit can reach approximately 380~km. Like Chariklo, the surrounding material at Chiron is thus very near or beyond the classical Roche limit. 

The locations of the significant material surrounding Chiron in 2018 are tens of km different from the proposed rings in 2011; however, they do fall within the errors of that ring-pole position. The optical depths in 2018 are lower than in 2011. Material around Chiron's nucleus has been previously observed while not in a ring-like configuration \citep[e.g.][]{RN1386, RN2024}. In addition, multiple thin and broad features were reported in 2011 by \citet{RN3716}. \citet{RN4001} most recently found that the structure of surrounding material changed between 2011 and 2022. The 2018 results presented here differ enough from the proposed two-ring system from 2011 of \citet{RN3627} to support the conclusion that the surrounding material at Chiron is evolving on relatively short timescales. More occultation observations are needed to continue studying Chiron and other intriguing and evolving small-body ring systems. 

\begin{acknowledgments}
We thank two anonymous reviewers for comments that improved the manuscript. This work was partially supported by National Science Foundation (NSF) Astronomy and Astrophysics Research Grants award number 2206306 and NASA grant 80NSSC21K0432. This paper uses observations made at the South African Astronomical Observatory (SAAO).  A portion this work was supported by the MIT-Germany University of Stuttgart Seed Fund (misti.mit.edu).  
\end{acknowledgments}

%

\vspace{5mm}
\facilities{SAAO}






\bibliography{Reference_Library_20230925}{}
\bibliographystyle{aasjournal}



\end{document}